# Who Plans, Who Edits? Verified, Cost-Aware Division of Labor for LLM-Driven Analog Layout Repair

Xilai Dai and Saeed Mohammadi

***Abstract*—Language-model agents can repair analog layouts when sign-off tools check every edit, but which model should plan a repair, which should edit it, and at what cost remain open questions. In our tool-grounded repair agent, models only choose bounded layout operators; sign-off tools rerun each task's contract checks after every edit, so failing layouts are never accepted. On layout-versus-schematic (LVS), design-rule checking (DRC), and post-layout repairs of three designs (a production high-voltage digital-to-analog converter (DAC), a 5-V oscillator, and a SKY130 operational transconductance amplifier (OTA)) with six tier-matched models from two vendors, 182 blind runs show that cheap models execute plans but rarely succeed alone (the cheapest model: 1 of 11 runs alone, 16 of 17 under a flagship planner; $p < 0.001$), cheap planners fail more often, and no fixed assignment is cheapest on every task. We propose a learning controller that selects models for each role and the escalation order from the first tool evaluation and updates Bayesian estimates after every attempt. With rules and initial estimates frozen before testing, the controller fixed 7 of 8 episodes on four sealed held-out tasks, matching the OpenAI flagship alone, at 15–40% of the flagship's spend per fix on three tasks but 8% higher spend overall. The cheapest mid-tier model alone fixed 6 of 8 at the lowest overall spend per fix but missed the hardest task. All 32 extracted DAC repairs match the human-guided LVS repair's integral nonlinearity (INL) within 0.005 least significant bits (LSB) at every code.**

## I. Introduction

An analog layout is finished only when three kinds of evidence agree: design-rule checking (DRC) accepts the geometry, layout-versus-schematic (LVS) checking accepts the connectivity, and simulation after parasitic extraction (PEX) shows that the circuit still meets its specification. When one of them fails, a designer diagnoses the cause, edits the layout, and reruns the tools. In an automated loop, most wall time is spent in the tools (Fig. 1(b)).

Large language model (LLM) agents can read verification reports and propose layout edits [1]–[4], and domain-adapted LLMs [5] and design-automation agents [6] have been demonstrated. Using them in practice raises a question unanswered by model routing [7]–[9], which assigns models per query: which model should perform each part of a tool-bound loop? Strong models are expensive per token; cheap ones fail more often. In a repair loop, a wrong decision costs full tool evaluations, while model spend is dominated by the context the models must read.

We study this question with a tool-grounded repair agent (Fig. 1) in which correctness never depends on the model. Models select operators from a bounded catalog; deterministic guards check and apply them to a copied layout; and native sign-off tools re-verify every edit, so a model can make a repair fail but cannot make a layout pass checks it fails. This turns the choice of models into a question of cost and time, which we answer in three steps. First, we separate a *planner*, which decides what to change, from an *executor*, which turns one intent at a time into operator calls. We measure tier-matched assignments of six models from two vendors to these roles. Second, an executor ladder repeats a failed attempt with a stronger executor. Third, because no fixed assignment is cheapest on every task, a learning controller chooses the assignment and the escalation order per task from the first tool evaluation and updates its estimates after every attempt.

We evaluate the agent on three designs that stress different evidence: LVS repair of a production 8-bit high-voltage digital-to-analog converter (DAC), a voltage-recognition DRC fix in a 5-V oscillator, and post-layout bandwidth and stability repair of a SKY130 operational transconductance amplifier (OTA), followed by four sealed held-out tasks. Our contributions are:

1) A verified repair agent with deterministic guards, an evidence builder that attributes post-layout degradation to parasitic groups and localizes design-rule markers, and a per-role ledger of tokens, spend, and tool time (Section IV).

2) A tier-matched, repeated measurement of who should plan and who should edit (182 runs): cheap models execute but rarely succeed alone, cheap planners fail more often, and interacting OTA repairs favor one strong model (Section VI-B).

3) A learning controller that assigns models to roles and orders escalation per task, with Bayesian estimates updated after every attempt and an optional LLM mapping for unseen task classes (Section IV-F).

4) A prospective evaluation on four sealed held-out tasks

*(Corresponding author: Xilai Dai.)*
X. Dai and S. Mohammadi are with the Elmore Family School of Electrical and Computer Engineering, Purdue University, West Lafayette, IN 47907, USA (e-mail: dai185@purdue.edu; saeedm@purdue.edu).

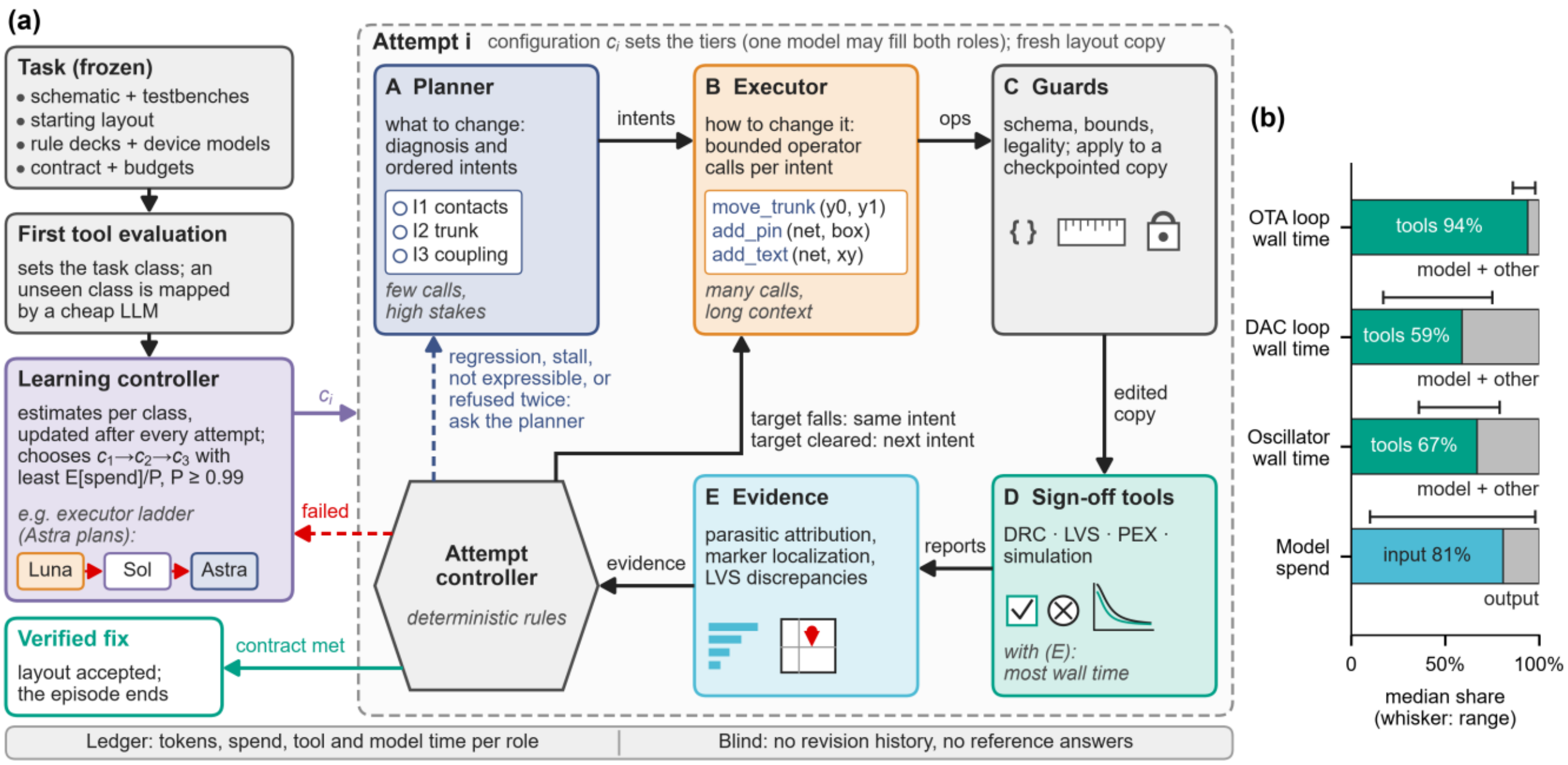

Fig. 1. Verified repair with a learned division of labor. (a) Left: the task's first tool evaluation sets its class; the learning controller (Section IV-F) chooses a sequence of configurations. Right: attempt $i$ runs configuration $c_i$ on a fresh copy of the starting layout. The planner sets intents (A); the executor turns each into operator calls (B); guards check and apply them to a checkpointed copy (C); sign-off tools re-verify every edit (D); the evidence adds attribution, marker localization, and LVS discrepancies (E). The attempt controller keeps the executor on an intent while its target falls and otherwise moves on or asks the planner (Algorithm 1). A met contract ends the episode with a verified fix; a failed attempt returns to the learning controller for the next configuration; every outcome updates the estimates. (b) Median shares (whiskers: range) of wall time in tool runs per task and of model spend on input context; 182 runs; 178 with model calls.

against three fixed baselines, with the controller's rules and initial estimates frozen beforehand, and production evidence: every extracted DAC repair behaves like the human-guided LVS repair, and DRC-count shortcuts through fixed foundry cells show that legality must be part of the contract (Sections VI-D to VI-F).

## II. Related Work

### A. Analog Layout Automation

Algorithmic systems such as ALIGN [10] and MAGICAL [11] generate placed and routed analog layouts, generator frameworks such as BAG2 [12] encode layout procedures, and learned guidance steers routing toward electrical goals [13]. These systems create layouts; we study the repair of an existing layout against a frozen schematic and a contract checked by sign-off tools.

### B. Language Models for Analog Design and Layout

Domain-adapted and agentic LLMs assist chip design [5], [6], generate analog circuits [14]–[16], and connect design intent to layout [17]; OSIRIS provides a scalable dataset-generation pipeline for machine learning in analog design [18]. LayoutCopilot coordinates agents for interactive analog layout [1], in-context policy improvement refines layouts from simulation feedback [2], and DRC-Aid and EvoDRC repair design-rule violations with agentic loops [3], [4]. These works report one model or one agent design; none asks how models with different prices should divide a tool-bound loop or what a verified fix costs.

### C. Multi-Model Systems and Adaptive Selection

Cascades and routers send easy queries to cheap models and hard ones to strong models [7]–[9], multi-LLM systems learn role assignments or agent architectures [19], [20], and planner/executor decompositions separate decision-making from execution [21]–[23]. Our controller differs in three ways: its signal is the verdict of sign-off tools, whereas cascades and routers act on model confidence or learned preferences; it assigns a model to each role of a task, while routers assign one model per query; and it learns from verified outcomes with Bayesian posteriors (with Thompson sampling [24], [25] in simulation). Agents that optimize a measurable proxy can satisfy it without achieving the goal [26]; Section VI-F reports such behavior for DRC counts.

## III. Repair Problem and Cost

A repair task specifies a schematic with its testbenches, a technology, a starting layout, a contract, and budgets. The contract has a structural part (no actionable DRC findings in scope, LVS match) and, where applicable, an electrical part stated as limits at every process/temperature condition. Every attempt starts from its own copy of the starting layout and sees only its own evidence. Budgets bound edits, model calls, tool evaluations, and wall time.

An attempt alternates among model decisions, edits, and tool evaluations. Its wall time is

$$W_{\text{att}} = \textstyle\sum_k t_{\text{tool},k} + \sum_j t_{\text{model},j} + t_{\text{ctrl}}, \quad (1)$$

where $k$ runs over tool evaluations, $j$ over model calls, and $t_{\text{ctrl}}$

is the attempt controller's own time. The model spend is

$$S = \sum_j [p_{\text{in}}(n_{\text{in}} - n_c) + p_c n_c + p_{\text{out}} n_{\text{out}}], \quad (2)$$

with input, cached-input, and output token counts $n$ and list prices $p$ of the model that served call $j$. A *verified fix* is an attempt whose final layout meets the contract on fresh tool evidence. Because failed attempts also incur costs, configurations and policies are compared by the spend per verified fix over all their attempts,

$$S_{\text{fix}} = \left(\sum_a S_a\right) / (\text{number of verified fixes}), \quad (3)$$

and likewise by the wall time per verified fix. Reliability enters as a constraint: a policy is preferred only if its estimated probability of reaching a verified fix within its escalation budget meets a target.

## IV. Verified Repair Agent

### A. Loop

The attempt controller (Fig. 1(a)) is deterministic code: it builds the evidence, chooses which model to call, checks and applies edits, runs the tools, and keeps the ledger. Model workers are memoryless and tool-free: each call receives one self-contained prompt and returns one JSON object; a malformed reply prompts one request to resend valid JSON. Every repair-model prompt is stored before the call, so the information available to each model can be audited.

### B. Evidence

The evidence packet is rebuilt from the tools after every edit. It contains the parsed DRC results with a bounded excerpt of the findings for every actionable rule, the LVS result with an excerpt of its discrepancy section, the electrical metrics at every condition, the pin voltages where a rule depends on them, and a compact layout model (DAC: shapes, vias, pins, and labels; oscillator: text labels; SKY130: routing trunks, tracks, and devices, with nets named from the candidate's own extraction once it matches LVS). *Parasitic attribution* runs when a layout is structurally clean but misses an electrical limit. Under the worst condition for each violated metric, it re-simulates the extracted netlist with the resistance or capacitance of one net, layer, or net pair removed. It reports each removal's share of the metric gap between the extracted netlist with and without all its parasitics. *Marker localization* gives the coordinates of the findings. In a hierarchical layout such as the DAC, it groups the findings by the instance that contains them, in the coordinates of the nearest editable cell.

### C. Operators and Guards

Models choose only operators and their parameters. The SKY130 catalog has eight geometry-derived operators (move a routing trunk, set a trunk or path width, maximize source/drain contacts, move a path endpoint, add or delete a via, relocate a tap). The DAC catalog has five (add a pin, rectangle, path, or via; delete a shape). The oscillator catalog has two (add or delete text on technology-defined layer/purpose pairs). Before any tool runs, deterministic guards check the schema, the bounds, and the editable cell. In the second version of the DAC's targeted-DRC extension (Section VI-F), a legality rule also refuses a drawn shape that, at any placement of its cell, overlaps a fixed foundry pad or I/O cell. Every edit, comprising one to three operator calls (up to six on the DAC), is applied to a checkpointed copy and rolled back if any operator fails.

### D. Planner and Executor

The planner receives the edit history and the evidence with the layout model reduced to structure. It returns a diagnosis (evidence, hypothesis, falsifier) and an ordered list of intents, each naming its target (violated metrics, rules, or LVS), the mechanism, the operators, the scope, and the expected effect. The executor receives one intent with the full evidence and the observed effect of its earlier edits for that intent. It returns operator calls or a statement that the intent cannot be expressed. It cannot change the target or stop the attempt. Procedure Attempt of Algorithm 1 specifies when the attempt controller returns to the planner: a *regression* (the edit breaks a check that passed, adds a violation of a new kind, or raises the overall violation score), a *stall* (the edit is applied but the target measure does not fall), an intent that cannot be expressed or is refused twice, or a plan whose intents are all cleared. In a single-model configuration, each call returns both the diagnosis and the operators, and the attempt stops when a call returns none.

**Algorithm 1** Repair episode with a learned assignment of models to roles

**procedure** Episode(task)
1: $k \leftarrow$ class(evidence(starting layout)) ▷ Section IV-F
2: $(c_1, \ldots, c_m) \leftarrow$ sequence with least $\mathrm{E}[S]/P$, $P \geq 0.99$, $m \leq 3$ ▷ (4)
3: **for** $i = 1, \ldots, m$ **do**
4: (ok, layout) $\leftarrow$ Attempt($c_i$, fresh copy of the starting layout)
5: update the estimates of $(c_i, k)$ with ok, spend, and wall time
6: **if** ok **then return** layout ▷ verified fix
7: **return** not fixed
**procedure** Attempt($c$, layout) ▷ $c$ sets planner and executor
8: $E \leftarrow$ evidence(layout); why $\leftarrow$ initial diagnosis
9: **while** contract not met **and** budgets remain **do**
10: **if** why $\neq \emptyset$ **then**
11: plan $\leftarrow$ planner($E$, why, edit history); why $\leftarrow \emptyset$; **if** plan $= \emptyset$ **then stop**
12: current intent $\leftarrow$ first intent of plan; clear its edits and refusals
13: ops $\leftarrow$ executor(current intent, $E$, earlier edits for it)
14: **if** ops $= \emptyset$ **then** why $\leftarrow$ not expressible; **continue**
15: **if** guards refuse or applying ops fails **then** why $\leftarrow$ refused if 2nd in a row; **continue**
16: $E' \leftarrow$ evidence(layout)
17: **if** the edit regresses (Section IV-D) **then** why $\leftarrow$ regression
18: **else if** target($E'$) $= 0 <$ target($E$) **then** next intent (why $\leftarrow$ done if none)
19: **else if** target($E'$) $\geq$ target($E$) **then** why $\leftarrow$ stall
20: $E \leftarrow E'$
21: **return** (contract met, layout)

*For a single-model c, one call per iteration replaces lines 10–14, lines 17–19 do not apply, and the attempt stops when a call returns no operators.*

### E. Escalation

An attempt that ends without a verified fix, because its budget is exhausted or a model stops it, is followed by a new attempt from the starting layout with the next configuration in the escalation sequence. Attempts are independent, and no edit from a failed attempt is inherited. The *executor ladder* keeps the planner and moves the executor from the cheap tier to the mid tier and then to the flagship tier. The learning controller generalizes this: its escalation sequence may change the planner or the executor, or switch to a single model.

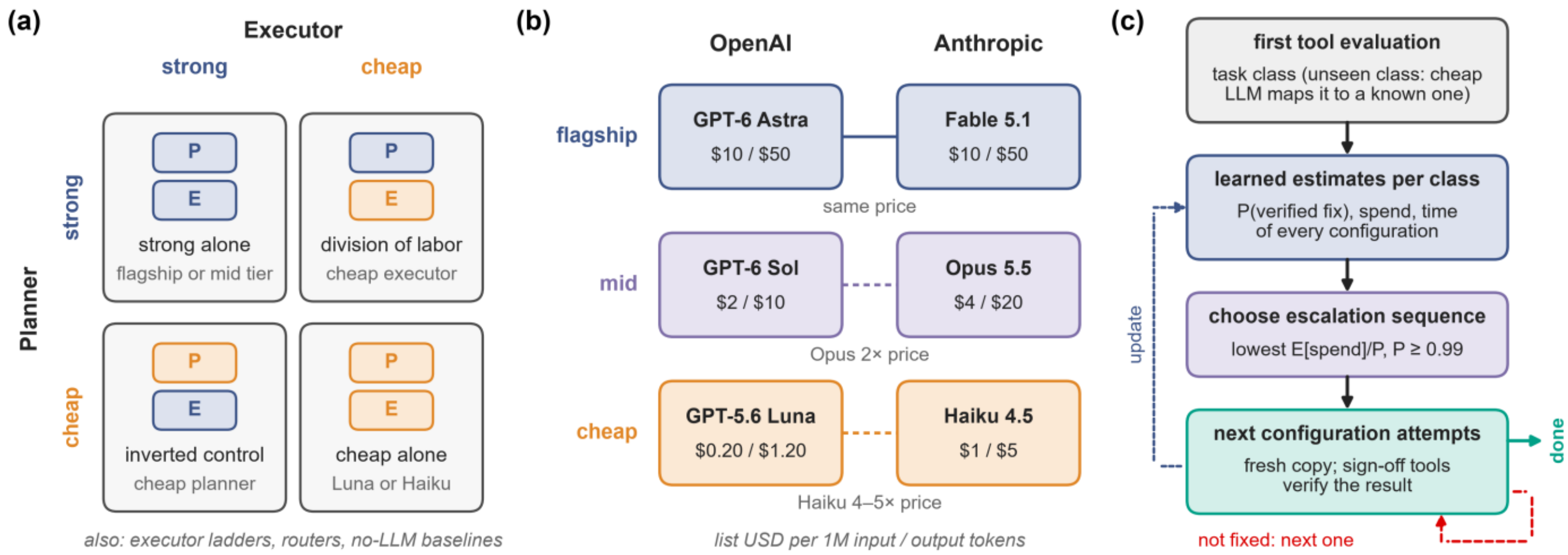

Fig. 2. Experimental design and controller. (a) Crossed planner/executor tiers; the inverted control swaps the roles of the Astra/Luna and Opus/Haiku pairs. (b) Tier-matched models of the two vendors with their list prices. (c) Learning controller: the first tool evaluation sets the task class (a cheap LLM maps an unseen class to a known one); learned estimates per class select an escalation sequence; attempts run from a fresh copy until one is verified or the sequence ends, and every outcome updates the estimates.

### F. Learning Controller

Before any repair-model call, the learning controller (Fig. 2(c)) computes a fingerprint of the task from its first tool evaluation and assigns it to a class: *connectivity* (LVS mismatch), *marker* (actionable DRC findings only), *electrical* (structurally clean, contract violated), or *mixed* (DRC findings and electrical violations). Its 20 configurations (Section V-B) are six single models, six planner/executor pairs, and the eight upper rungs of four executor ladders. For every configuration $c$ and class $k$, it maintains a Beta posterior on the probability of a verified fix and running means of spend and wall time per attempt. The prior has a weight of two observations, and its mean averages the configuration's results on other classes with a tier-based guess (fix probability 0.35, 0.6, or 0.8 for a cheap, mid, or flagship model alone, lower for pairs). Thus, experience is shared across classes but overridden by data from that class. For an escalation sequence $(c_1, \ldots, c_m)$, $m \leq 3$, of independent attempts with success probabilities $p_i$ and mean spends $s_i$, the expected spend and the probability of a fix are

$$\mathrm{E}[S] = \textstyle\sum_i \; s_i \prod_{l<i}(1 - p_l)\,, \quad P = 1 - \prod_i \; (1 - p_i). \quad (4)$$

The learning controller selects, by exhaustive search, the sequence with the lowest $\mathrm{E}[S]/P$ among those with $P \geq 0.99$ (the highest $P$ if none reaches it), breaking ties within 10% by the expected wall time per fix. Each attempt is verified, so the tools enforce correctness; the 0.99 target only sets how much escalation to budget. After every attempt, whether successful or not, the posterior of its configuration and class is updated (procedure EPISODE in Algorithm 1). In the simulated cold start of Section VI-D, decisions sample from the posteriors (Thompson sampling [24], [25]); all other decisions, including those on the held-out tasks, use the posterior means. When a task's class is unseen, an optional cheap LLM receives the tool fingerprints of the new task and of the known classes, never the layout, and names the most similar known class. In the held-out runs, the LLM was needed once and cost $0.002.

## V. Experimental Design

### A. Tasks

*DAC (connectivity).* A production 8-bit DAC with a 20-V output range in a commercial 0.18-µm high-voltage bipolar-CMOS-DMOS (BCD) process. Its original layout fails LVS; the agent must reach LVS match without editing the schematic, with LVS, not DRC, rerun after every edit (budget: six edits, ten calls, 2 h). Two human-guided layouts, r01 (LVS repair) and r14 (final), are references never shown to the agents.

*Oscillator (marker).* A 5-V ring oscillator in the same process whose layout matches LVS but has 33 findings of a voltage-recognition rule on the metal-1 strips of its capacitors. The agent must clear them without increasing the number of findings of any other rule and maintain LVS match, with whole-chip DRC and LVS rerun after every edit (budget: six edits, ten calls). The evidence includes the rule description, the markers, and the design's pin voltages.

*OTA (electrical).* A two-stage fully differential OTA with common-mode feedback in SKY130 [27], using release 0.1.0 of its Cadence OpenAccess process design kit (PDK). The OTA's layout matches LVS but loses 12.3–20.8% of its unity-gain bandwidth (UGB), and its common-mode phase margin (CM PM) falls to 11.6°. The contract requires no actionable DRC findings (density rules are reported only), LVS match, a UGB loss of at most 10%, and a CM PM of at least 30° at nine conditions (process corners TT, SS, and FF at 0, 27, and 70 °C). Each evaluation runs DRC and LVS and, once LVS matches, extraction and nine simulations (budget: four edits, eight calls, five evaluations).

*Held-out tasks.* Four tasks on a separate SKY130 OTA instance with PDK resistors and metal–insulator–metal (MIM)

TABLE I
MODELS AND TIERS

| Tier | OpenAI model | Price ($/M in / out) | Anthropic model | Price ($/M in / out) |
|---|---|---|---|---|
| flagship | GPT-6 Astra | 10 / 50 | Claude Fable 5.1 | 10 / 50 |
| mid | GPT-6 Sol | 2 / 10 | Claude Opus 5.5 | 4 / 20 |
| cheap | GPT-5.6 Luna | 0.20 / 1.20 | Claude Haiku 4.5 | 1 / 5 |

List API prices; spend is computed from logged tokens at these prices, with cached input at the full input price for the OpenAI models (Section VII); the Anthropic workers reported no cached input. All models ran through tool-free command-line workers at low reasoning effort, except Haiku, whose worker exposes no effort setting.

TABLE II
CONFIGURATIONS

| Kind | Configurations | Role in the study |
|---|---|---|
| single model | each of the six models alone | references at every tier |
| strong plans, cheap edits | Astra/Luna, Fable/Haiku, Sol/Luna, Opus/Haiku | division of labor |
| cheap plans, strong edits | Luna/Astra, Haiku/Opus | inverted control |
| executor ladder | planner fixed; executor cheap→mid→flagship | escalation |
| learning controller | chosen per task (Section IV-F) | proposed method |
| routers, ablation, no-LLM | Luna/Astra and Haiku/Opus routers; Astra without attribution; enumeration, random | baselines (OTA) |

capacitors use the same operators and contract. They were prepared and sealed before any agent ran on them: a metal-width DRC fault (A17), an open connection (B29), degraded source/drain contacts that cost 17–24% of the UGB (C44), and a combination of DRC and contact faults (D58). Their task cards contain no fault label or solution.

### B. Models and Configurations

Table I and Fig. 2(b) list the six models, two per tier; the tiers order each vendor's models by list price. The flagship models have the same price, Opus costs twice as much as Sol, and Haiku costs four to five times as much as Luna. Table II and Fig. 2(a) give the configurations built from them.

### C. Protocol

All repair agents worked under blind conditions: each attempt ran in its own sandboxed copy of the design and saw only the evidence from its own tool runs, never a reference solution, a revision history, or the held-out answer key. Every single-model and planner/executor configuration was run at least five times on the DAC and three times on the oscillator and the OTA. Each executor ladder was run for five episodes on the DAC and three on the other tasks. The first attempt of a ladder episode is counted as a run of its planner/executor pair. In all, the three calibration tasks comprise 182 runs: 169 attempts of the learning controller's configurations (25 of them upper ladder rungs) and 13 OTA runs of routers, an ablation, and no-LLM baselines. The learning controller's decision code, initial state, and update rule were frozen and hashed before the first held-out run. The controller then ran two rounds of one episode per held-out task. Episodes within a round could overlap, and each chose its sequence from the experience logged at its start. The learning controller continued to update its estimates after every attempt, as in deployment. Three fixed baselines ran two episodes per task: Sol alone (the cheapest mid-tier model) and Astra alone (the OpenAI flagship), each with one attempt per episode, and the executor ladder with Astra planning. Runs invalidated by an infrastructure fault, such as a full disk or a defect in parsing a tool report, were discarded, removed from the controller's log, and repeated with the same configuration. Their spend is not included in any reported figure.

### D. Measurements and Statistics

The ledger records every model call (role, model, tokens, latency) and every tool run. Success rates are given with 95% Wilson intervals [28] and spend per verified fix is reported with 95% percentile bootstrap intervals over runs [29] (unbounded when more than 2.5% of resamples have no verified fix; approximate with three runs). Contrasts use two-sided Fisher exact tests without correction for multiple comparisons. DAC post-layout validation compares integral and differential nonlinearity (INL, DNL), in least significant bits (LSB), after full RC extraction with coupling capacitance and a transient simulation over all 256 codes (0.2-pF load, 27 °C, 10-µs dwell). The simulator's post-layout RC-reduction option is disabled because it removes the input nets of the DAC, which reach only transistor gates through extracted resistors.

## VI. RESULTS

### A. Where Time and Money Go

Fig. 1(b) summarizes the ledger over 182 runs. Tool runs took a median of 94% of wall time on the OTA (one evaluation includes an extraction, nine simulations, and any attribution runs), 59% on the DAC (a whole-chip LVS run), and 67% on the oscillator (a whole-chip DRC and LVS run). Over the 178 runs that called a model, input context accounted for a median of 81% of model spend, and input tokens outnumbered output tokens by a median of 22:1. Wall time therefore falls mainly through fewer evaluations, and spend falls mainly through fewer, cheaper, or shorter-context calls.

### B. Who Plans, Who Edits

Table III and Fig. 3 give the outcomes. Every strong model alone, at the flagship or the mid tier, fixed every run of every task. On every task, the cheapest verified option was a mid-tier model alone, never a pair: Sol on the DAC ($0.28 per fix) and the OTA ($0.11), Opus on the oscillator ($0.07). Four findings

TABLE III
VERIFIED FIXES AND SPEND PER VERIFIED FIX ON THE CALIBRATION TASKS

| Configuration | DAC fixes | DAC $/fix [95% CI] | OTA fixes | OTA $/fix [95% CI] | Osc. fixes | Osc. $/fix [95% CI] |
|---|---|---|---|---|---|---|
| Astra alone | 5/5 | $1.19 [1.11, 1.34] | 3/3 | $0.27 [0.27, 0.27] | 3/3 | $0.19 [0.19, 0.19] |
| Fable alone | 5/5 | $0.91 [0.77, 1.05] | 3/3 | $0.55 [0.52, 0.60] | 3/3 | $0.20 [0.12, 0.26] |
| Sol alone | 5/5 | $0.28 [0.22, 0.36] | 3/3 | $0.11 [0.06, 0.17] | 3/3 | $0.11 [0.07, 0.15] |
| Opus alone | 5/5 | $0.42 [0.39, 0.48] | 3/3 | $0.32 [0.28, 0.40] | 3/3 | $0.07 [0.05, 0.09] |
| Luna alone | 0/5 | — | 0/3 | — | 1/3 | $0.08 [0.02, ∞) |
| Haiku alone | 1/5 | $2.58 [0.69, ∞) | 0/3 | — | 0/3 | — |
| Astra plans, Luna edits | 9/10 | $0.58 [0.54, 0.67] | 4/4 | $0.43 [0.25, 0.61] | 3/3 | $0.58 [0.57, 0.58] |
| Fable plans, Haiku edits | 2/5 | $2.54 [0.97, ∞) | 0/3 | — | 3/3 | $0.28 [0.14, 0.54] |
| Sol plans, Luna edits | 3/5 | $0.36 [0.15, 1.37] | 1/3 | $0.44 [0.11, ∞) | 1/3 | $0.26 [0.04, ∞) |
| Opus plans, Haiku edits | 7/10 | $0.72 [0.47, 1.41] | 0/4 | — | 3/3 | $0.14 [0.08, 0.20] |
| Luna plans, Astra edits | 3/5 | $2.18 [1.44, 8.03] | 0/3 | — | 0/3 | — |
| Haiku plans, Opus edits | 4/5 | $0.99 [0.67, 2.35] | 0/3 | — | 2/3 | $0.40 [0.11, ∞) |
| executor ladders (episodes) | 20/20 | | 7/12 | | 12/12 | |

Fixes: verified fixes / runs. Spend per verified fix includes the spend of failed runs; "—" where no run of the configuration was verified; "[$a$, ∞)": unbounded interval (Section V-D). Planner/executor rows pool dedicated runs with the first attempts of ladder episodes; the ladder row counts episodes fixed at any rung, without spend per fix.

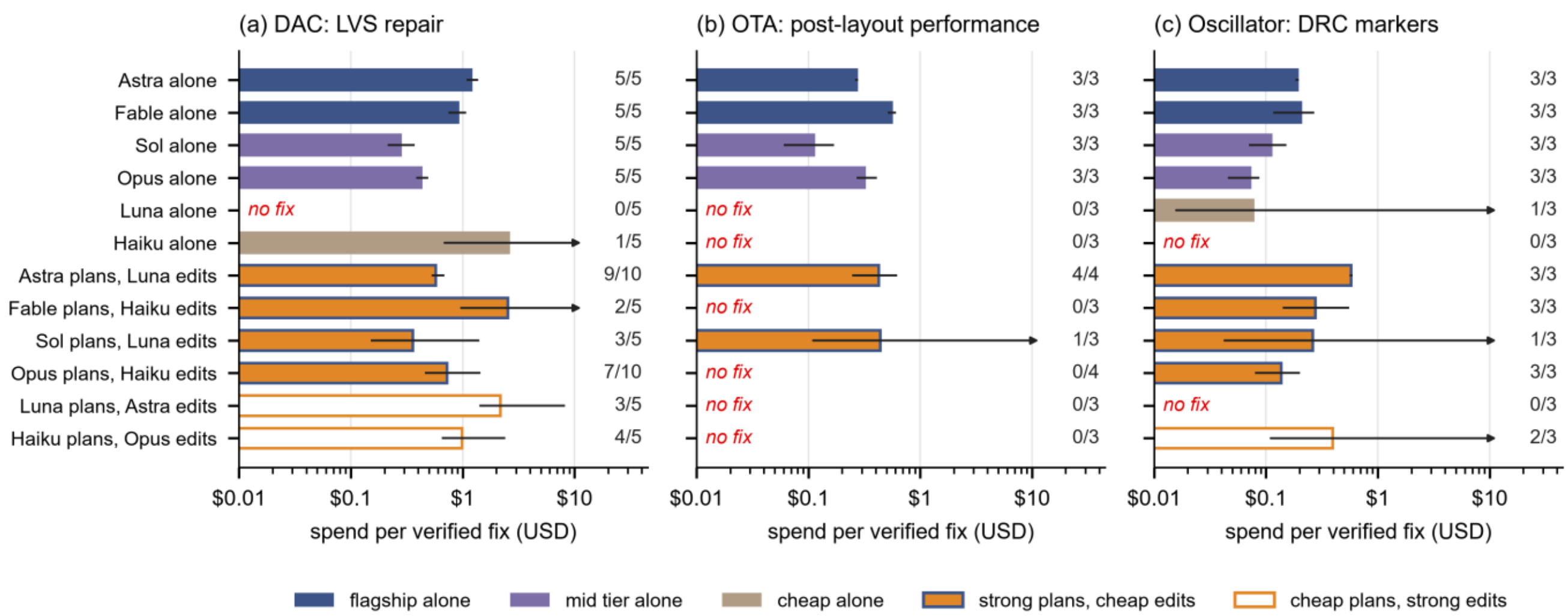

Fig. 3. Spend per verified fix of each configuration on the three calibration tasks, (a) DAC, (b) OTA, and (c) oscillator (bars; whiskers: 95% bootstrap interval; arrow: unbounded), with verified fixes per run at the right.

follow; Table IV tests the main contrasts.

*Cheap models execute but rarely succeed alone.* Alone, Luna and Haiku each fixed 1 of 11 runs; as executors for Astra, and for Opus or Fable, they fixed 16 of 17 and 15 of 28 (Luna for Sol: 5 of 11). A planner that decides what to change makes the cheap models usable editors and makes Luna a reliable editor under Astra. Without the planner, cheap models chose wrong text layers, stopped early, or exhausted their edits.

*The planner is the role that needs capability.* With the same two models and the roles swapped, the cheap planner succeeded in 9 of 22 runs against 26 of 34 for the strong planner, a gap carried by Luna and Astra (3 of 11 vs. 16 of 17; Haiku and Opus: 6 of 11 vs. 10 of 17). Some cheap plans could not be used at all: in all three oscillator runs, Luna as planner returned malformed JSON twice in a row.

TABLE IV
KEY CONTRASTS (CALIBRATION RUNS)

| Contrast | Fixes | 95% CI (%) | $p$ |
|---|---|---|---|
| Luna: alone vs. for Astra | 1/11 vs. 16/17 | 2–38 vs. 73–99 | <0.001 |
| Haiku: alone vs. for Opus, Fable | 1/11 vs. 15/28 | 2–38 vs. 36–70 | 0.014 |
| planner: cheap vs. strong | 9/22 vs. 26/34 | 23–61 vs. 60–88 | 0.011 |
| OTA: Sol, Opus, Fable plan vs. alone | 1/10 vs. 9/9 | 2–40 vs. 70–100 | <0.001 |

Fix rates with 95% Wilson intervals; two-sided Fisher exact tests.

*Dividing the work rarely pays.* With Astra planning and Luna editing, the spend per fix on the DAC was $0.58 against $1.19 for Astra alone. As planner, Astra was consulted about twice per run on the reduced evidence and used 39% as many of its own tokens as when working alone. Only this pair fixed every OTA run (4/4), though at $0.43 per fix against $0.27 for Astra alone. With Opus or Fable planning, Haiku as executor was not cheap enough per token, wrote long replies, and often failed, so

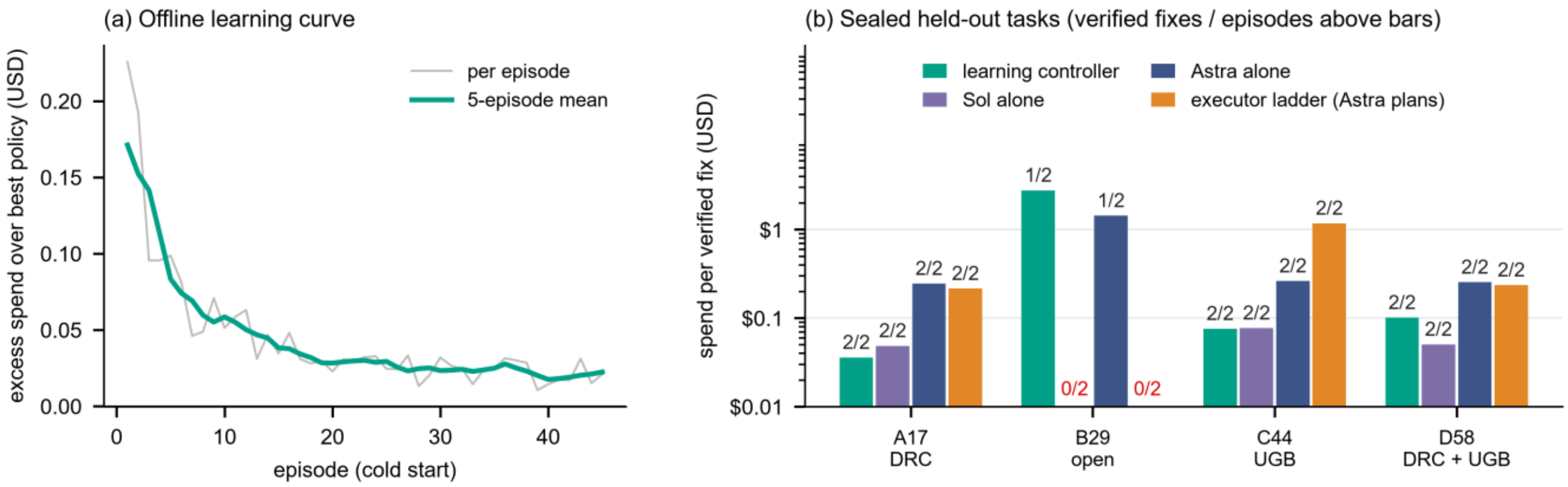

Fig. 4. Learning controller. (a) Offline learning curve: spend of each episode minus the expected spend per fix of the hindsight-best policy for its task, averaged over 200 cold starts in random task order with outcomes resampled from the measured runs (Thompson sampling over the 13 configurations measured on every task, without the wall-time tie-break and with the most probable single configuration as fallback). (b) Sealed held-out tasks: spend per verified fix of the controller, Sol alone, Astra alone, and the executor ladder with Astra planning, with verified fixes per two episodes above the bars.

TABLE V
SEALED HELD-OUT TASKS (TWO EPISODES PER TASK AND METHOD)

| Task | Controller fixes | Controller $/fix | Sol fixes | Sol $/fix | Astra fixes | Astra $/fix | Ladder fixes | Ladder $/fix |
|---|---|---|---|---|---|---|---|---|
| A17 (DRC) | 2/2 | $0.036 | 2/2 | $0.049 | 2/2 | $0.245 | 2/2 | $0.215 |
| B29 (open) | 1/2 | $2.771 | 0/2 | — | 1/2 | $1.431 | 0/2 | — |
| C44 (UGB) | 2/2 | $0.076 | 2/2 | $0.077 | 2/2 | $0.262 | 2/2 | $1.176 |
| D58 (DRC + UGB) | 2/2 | $0.101 | 2/2 | $0.050 | 2/2 | $0.254 | 2/2 | $0.236 |
| All | 7/8 | $0.457 | 6/8 | $0.099 | 7/8 | $0.422 | 6/8 | $1.016 |

Spend per verified fix over both episodes, including failed attempts and the controller's class mapping; "—": no verified fix. One attempt per episode for Sol and Astra. Ladder: Astra planning; executor Luna→Sol→Astra.

the divided runs cost more per fix than the planner alone.

*Interacting performance repairs favor one strong model.* On the OTA, the two failures interact: the ground network's resistance costs bandwidth and the coupling between two trunks costs CM PM, and a fix for one can disturb the other. Alone, Sol, Opus, and Fable fixed 9 of 9 runs. As planners handing intents to a cheap executor, however, they fixed 1 of 10; as later ladder rungs executing their own plans, they fixed only 1 of 7. Splitting the repair into intents limits these planners even with a strong executor, although Astra's plans fixed 4 of 4 with Luna. One model that sees all the evidence can make a single joint fix.

*Routers and baselines.* On the OTA, rule-based routers send a decision to the strong model at the start of a multi-metric repair or after a refused or non-improving edit, and otherwise to the cheap one. These routers fixed 3 of 3 with Luna and Astra (every decision went to Astra, at $0.27 per fix) and 0 of 3 with Haiku and Opus. Without parasitic attribution, Astra fixed 2 of 3 at $1.15 per fix, about four times its spend with attribution. Enumeration (deterministic, one run) and random search (three runs) over single operations of three operator types (contacts, trunk width, trunk position), with greedy acceptance within the same five-evaluation budget, fixed none.

## C. Escalation

The executor ladders fixed every episode on the DAC (20/20) and the oscillator (12/12) but only 7/12 on the OTA. There, the ladder with Astra planning fixed all its episodes on the first attempt and those with Fable, Sol, and Opus planning fixed 2, 1, and 1 of 3. On the OTA, executor escalation recovered only 3 of the 8 episodes whose first attempt failed.

## D. Learning Controller

*Offline.* Given the runs from two calibration tasks, the controller made its selection for the third, whose class it had never seen. It chose among all 20 configurations or one vendor's 10 (three tasks × both vendors, OpenAI only, Anthropic only) without the LLM mapping. In 7 of 9 cases its sequence was as cheap per fix as the cheapest policy that never failed on the third task's measured runs. In the other two, it cost at most 1.55 times as much (configurations not evaluated on the third task are skipped when a sequence is scored). Its first choice was always a mid-tier model alone, so the fixed rule "cheapest mid-tier model alone" scores the same here. In a simulated cold start with Thompson sampling over the 13 configurations measured on all three tasks, its mean excess spend per episode fell from $0.101 over the first ten episodes to $0.022 over the last ten (Fig. 4(a)).

*Prospective.* Starting from 169 calibration attempts using 18 of its 20 configurations, the controller chose single models, cheapest first, for the marker class (Luna, then Sol, then Opus). For the connectivity and electrical classes, it chose Sol alone, then Astra planning with Luna editing, then Astra alone. For the mixed task, whose class it had not seen, it followed the LLM mapping to the electrical class in the first episode and used the mixed class's own experience in the second. After the first B29 episode required all three attempts, it replaced the pair with Opus alone in the second. Table V and Fig. 4(b) give the outcomes. The controller fixed 7 of 8 episodes, as many as Astra alone and more than Sol alone and the ladder (6 of 8 each). On A17, C44, and D58 its spend per fix was 15–40% of Astra's. The B29 open-connection task is the exception, and its difficulty is a property of the evidence rather than of any assignment. The LVS report names the open net. However, until LVS matches, the layout model has no net names, so a fix requires inferring which wire carries the net. The executors refused to guess and the ladder fixed neither episode; Sol alone stopped in one episode and exhausted its budget in the other; Astra alone and the controller each fixed one. There, the controller's first choice, learned on the DAC, did not transfer. Its escalation incurred higher spend than Astra alone, so its overall spend per fix ($0.457) was 8% above Astra's ($0.422). The geometric mean of its per-task spend ratios to Astra was 0.42. The Sol-alone baseline, also the controller's first choice for the connectivity and electrical classes, fixed every episode of A17, C44, and D58 and spent $0.099 per verified fix overall, 23% of Astra's.

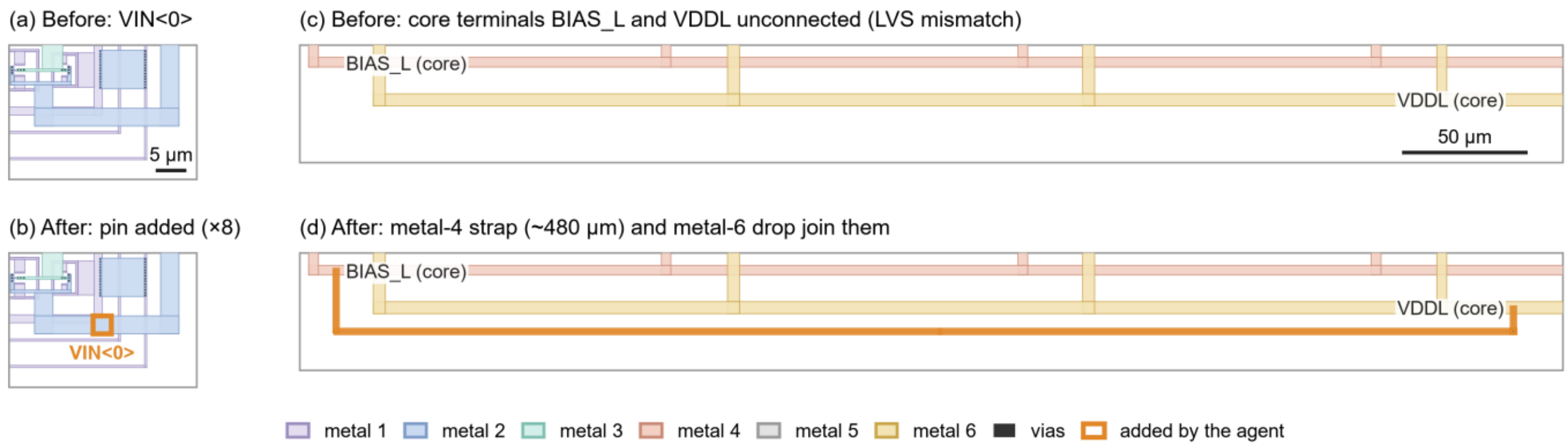

Fig. 5. Agent repair of the DAC connectivity faults at the top level; shapes added by the agent are outlined. (a), (b) One of the eight digital inputs before and after a pin and label were placed over the core's input terminal. (c), (d) The core's BIAS_L and VDDL terminals before and after being joined by a metal-4 strap and a metal-6 drop. Layers are identified by metal level only.

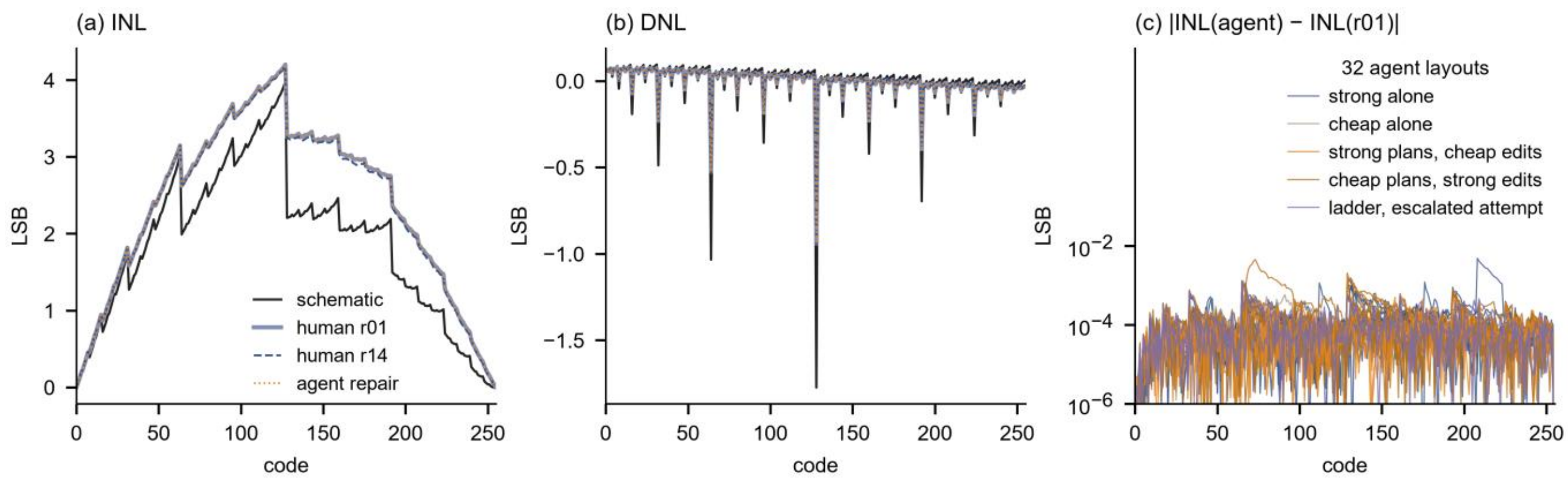

Fig. 6. DAC behavior over all 256 codes after full RC extraction. (a) Endpoint INL and (b) DNL of the schematic, the human-guided layouts r01 and r14, and an agent repair (layout curves overlap). (c) Absolute per-code INL difference of every extracted agent repair from r01 (log scale).

Cheap-first escalation also costs time: the controller's wall time per fix was 1018 s against 548 s for Astra alone and 741 s for Sol alone. This additional time arose because failed cheap attempts require tool evaluations before the next attempt starts and two of the controller's Sol fixes (the first episodes of C44 and D58) took two edits whereas Astra's took one. The three fixed choices thus mark the trade-off: Sol alone is cheapest but misses the hard task, Astra alone is fastest, and verified escalation matches Astra's fixes at lower spend than Astra on the three tasks Sol also fixes.

### *E. Correctness and Post-Layout Behavior*

No accepted layout failed its contract. Every failure ended with the contract unmet on fresh tool evidence. We re-checked the saved final evidence of all 51 accepted OTA layouts, both calibration and held-out: each shows LVS match, no actionable DRC findings, and finite metrics within the limits at all nine conditions. The OTA acceptance check required nine evaluated conditions but not every metric at each; the re-check found none missing. On the DAC, every verified repair fixed both connectivity faults of the original layout (eight digital inputs without top-level pins; the core's BIAS_L disconnected from VDDL) and reached LVS match in 2 to 5 edits (Fig. 5).

Table VI and Fig. 6 compare post-layout behavior. We extracted and simulated 32 of the 55 verified DAC repairs, including at least one from each of the 14 configurations that produced one (11 in Table III and three upper ladder rungs). Their maximum INL spans 4.2016–4.2018 LSB against 4.2016 for r01. Their maximum DNL spans 0.9424–0.9426 against 0.9425. The INL of every code differs from r01 by at most 0.005 LSB. Their maximum INL and DNL are within 0.016 and

TABLE VI
DAC POST-LAYOUT BEHAVIOR OVER 256 CODES

| Layout | Max \|INL\| (LSB) | Max \|DNL\| (LSB) | FS err. (mV) | Power (mW) |
|---|---|---|---|---|
| Schematic | 3.98 | 1.77 | −1.95 | 5.50 |
| Human-guided r01 (LVS repair) | 4.20 | 0.94 | +1.54 | 5.71 |
| Human-guided r14 (final) | 4.19 | 0.95 | −0.54 | 5.71 |
| Agent repairs, min–max (32 layouts) | 4.20 | 0.94 | +1.53 to +1.54 | 5.71 |

FS err.: full-scale error, code-255 output minus 255 × 78.125 mV. INL/DNL use each layout's own endpoint LSB (≈78.1 mV). The schematic has two negative code steps; no extracted layout has any.

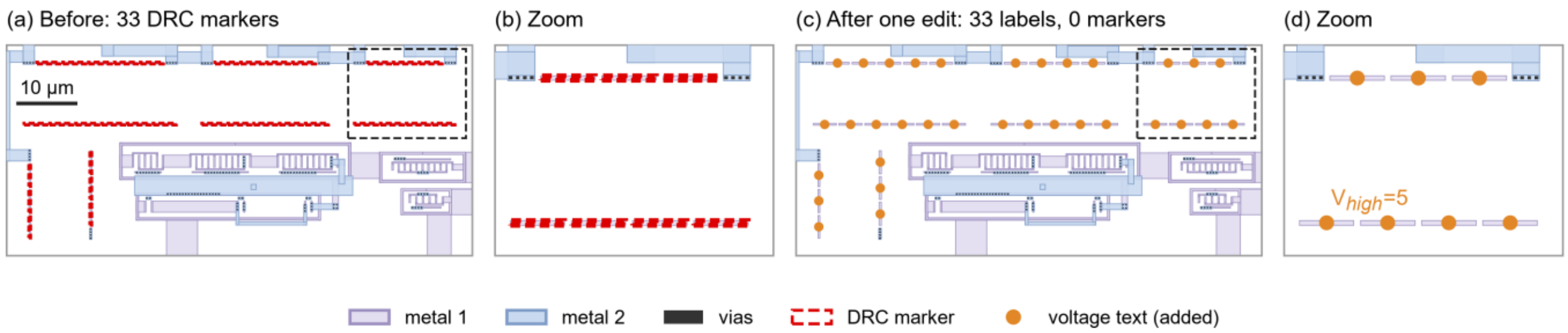

Fig. 7. Oscillator repair, partial view. (a) The 33 findings of the voltage-recognition rule (red) mark metal-1 capacitor strips whose nets carry no voltage text; (b) zoom of the dashed box. (c) After one agent edit, a voltage label on every flagged strip and no remaining finding; (d) zoom of the dashed box. LVS stays matched and no other rule changes.

0.0051 LSB of r14's. Re-extracting and re-simulating three of them reproduced every number exactly.

On the oscillator (Fig. 7), the repairs add text that informs the checker of each net's voltage range. Such text does not enter extraction. For a repair that labels all 33 strips, the extracted netlist matches the original in every resistor, capacitor, and device (only cell names and the date differ). A 120-point post-layout simulation (five process corners, three temperatures, eight control codes) gave identical frequency and power to four significant digits. On the OTA, one of Astra's one-edit repairs (Fig. 8) moved the common-mode drive trunk and the output trunk to quiet tracks and maximized source/drain contacts, giving a UGB loss of 7.8–9.5% and a CM PM of at least 42.6° at all nine conditions.

### *F. Legality Under Design-Rule Pressure*

In a targeted-DRC extension of the DAC task, whose targets were the high-voltage findings removed by the human-guided revisions, six of the first seven configurations tried, plus the first planner/executor run, drew shapes over fixed foundry pad cells to satisfy a salicide-block rule for I/O electrostatic-discharge (ESD) devices. One covered the pad region with blocks of 21 275 and 32 400 µm² and reduced that rule's findings from 122 to 20. Such edits modify qualified foundry cells from outside. In a second version with the legality guard of Section IV-C and that salicide-block rule reported rather than required, no configuration attempted such edits and none cleared the remaining high-voltage marker rules. The human-guided revisions cleared them only by moving instances, which no operator can do. A DRC count is a proxy; legality must be part of the contract.

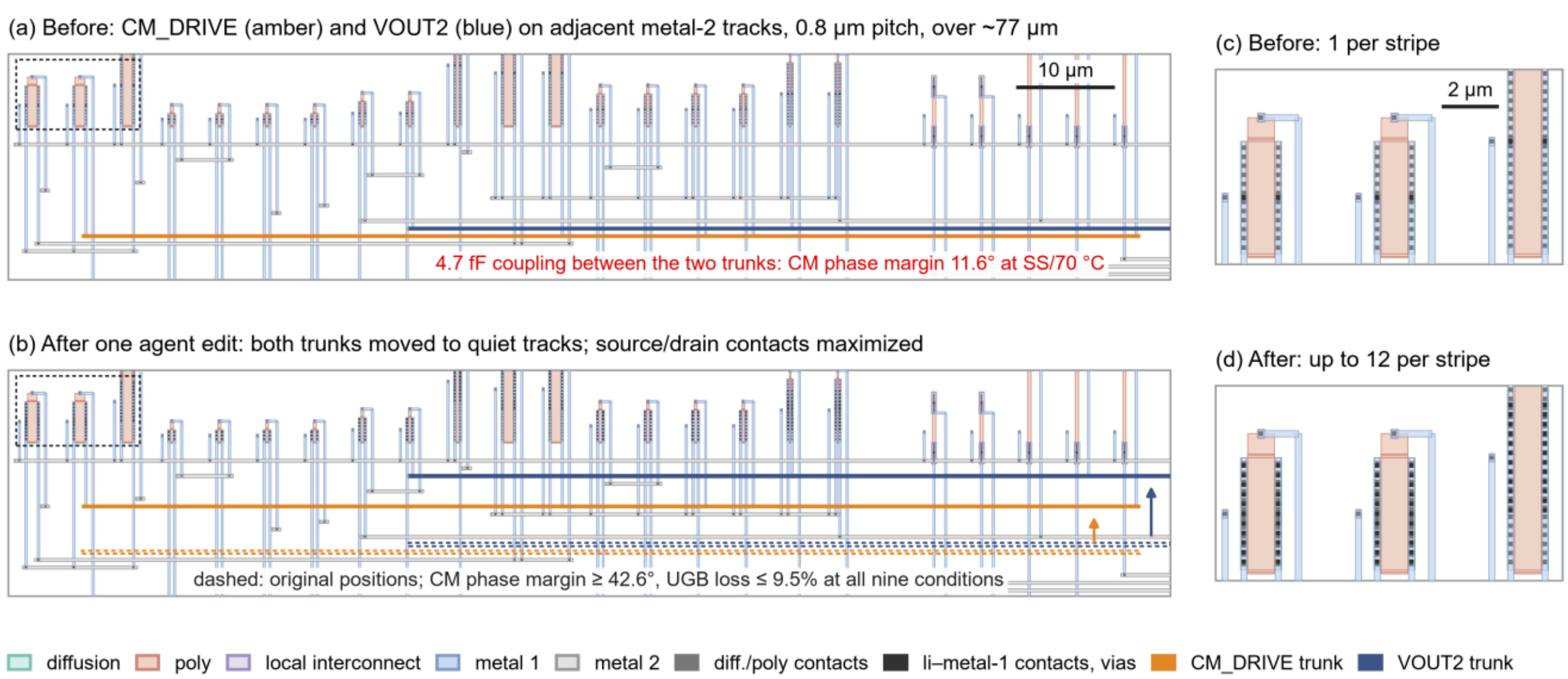

Fig. 8. OTA layout before and after a one-edit repair. (a) The common-mode drive trunk (amber) and the second-stage output trunk (blue) run on adjacent metal-2 tracks at a 0.8-µm pitch over about 77 µm; their 4.7-fF coupling lowers the CM phase margin to 11.6° at SS/70 °C. (b) Both trunks moved to quiet tracks (dashed: original positions) and source/drain contacts maximized. (c), (d) Local-interconnect-to-metal-1 contacts on the boxed device row before and after.

## VII. Discussion and Limitations

*What decides the assignment.* The data suggest three regimes. When the fix is local and its location is identified in the evidence (DRC markers, a clear bandwidth cause), a mid-tier single model suffices; a cheap one tried first can cost less. When the repair requires several independent edits and a cheap executor is much cheaper per token, terse, and reliable, a strong planner with a cheap executor costs less than the planner alone, as Astra and Luna did on the DAC, although Sol alone cost even less. When failures interact, or the evidence does not identify where to edit, a single strong model that sees everything is the safer choice. The controller learns per task class what is cheapest and escalates on failure.

*Evidence limits achievable repair performance.* When the evidence does not locate the object to edit, as on B29 (Section VI-D), executors that must take coordinates from the evidence refuse. Richer evidence, such as net names traced from the LVS database for a mismatched layout, could raise this ceiling for every configuration.

*Limitations.* The study covers three designs for calibration and one circuit for the held-out tasks, with three to ten runs per single-model or planner/executor configuration (fewer for later ladder rungs) and two held-out episodes per method. The held-out estimates are therefore uncertain: a 7/8 fix rate has a 95% Wilson interval of 53–98%, and the controller's advantage over Sol alone rests on one task. The controller's classes are coarse, its calibrated estimates adapt slowly when a class behaves differently on a new design, and (4) treats attempts as independent, so on B29 it predicted $P > 0.999$ for a task on which one of two episodes failed. Spend is calculated using list prices for subscription-served models; in the calibration runs the OpenAI workers reported 23–47% of their input as cached, which we charge at the full input price. At one tenth of the input price for cached tokens, held-out spend per fix would be \$0.324 for the controller, \$0.077 for Sol alone, \$0.269 for Astra alone, and \$0.685 for the ladder. On A17, C44, and D58 the controller's spend per fix would be 20–36% of Astra's. Tiers are defined by list price within each vendor, and the cheap tiers of the two vendors could not be matched in price.

## VIII. Conclusion

When every edit in an analog layout repair is verified by sign-off tools, the tools decide correctness, and choosing which language model plans and which edits becomes a question of cost and time. Over 182 blind runs (agents never saw reference layouts or held-out answers) on three designs, cheap models execute plans but rarely succeed alone, cheap planners fail more often than strong ones, and interacting OTA repairs favor one strong model. No fixed assignment is cheapest everywhere. A controller that learns the assignment per task class from verified outcomes, with its rules and initial estimates frozen before testing, matched the OpenAI flagship's verified fixes on four sealed held-out tasks at 15–40% of the flagship's spend per fix on three tasks, though with more time and slightly more spend overall. The cheapest mid-tier model alone spent the least but missed the hardest task. Every extracted repair of a production DAC behaves like the human-guided LVS repair.

## Acknowledgment

The authors used OpenAI Codex for grammar correction and language editing of the manuscript. The authors reviewed the edits and take responsibility for the final content.